\documentclass[10pt,letterpaper]{article}
\usepackage[top=0.85in,left=1in,footskip=0.75in,marginparwidth=1in]{geometry}

\usepackage[utf8]{inputenc}

\usepackage{cite}

\usepackage{amssymb}
\usepackage{latexsym}

\usepackage{url}
\usepackage{xcolor}
\usepackage{svg}
\usepackage{amsmath}
\usepackage{mathtools}
\usepackage{soul}
\usepackage{hyperref}
\definecolor{newcolor}{rgb}{.8,.349,.1}

\usepackage{microtype}
\DisableLigatures[f]{encoding = *, family = * }

\raggedright
\usepackage{changepage}

\usepackage[aboveskip=1pt,labelfont=bf,labelsep=period,singlelinecheck=off]{caption}

\makeatletter
\renewcommand{\@biblabel}[1]{\quad#1.}
\makeatother

\usepackage{lastpage,fancyhdr,graphicx}
\usepackage{epstopdf}
\fancyheadoffset[L]{2.25in}
\fancyfootoffset[L]{2.25in}

\usepackage{color}

\definecolor{Gray}{gray}{.25}

\usepackage{graphicx}

\usepackage{sidecap}

\usepackage{wrapfig}
\usepackage[pscoord]{eso-pic}
\usepackage[fulladjust]{marginnote}
\reversemarginpar

\begin{document}
\vspace*{0.35in}

% title goes here:
\begin{flushleft}
{\Large
\textbf\newline{DeLTA-BIT: an open-source probabilistic tractography-based deep learning framework for thalamic targeting in treatments of functional neurological disorder}
}
\newline
% authors go here:
\\

Mattia Romeo\textsuperscript{1,2,3,$\dagger$},
Cesare Gagliardo\textsuperscript{4,5,$\dagger$},
Grazia Cottone\textsuperscript{1,2,$\dagger$},
Giorgio Collura\textsuperscript{1,5},
Enrico Maggio\textsuperscript{1},
Claudio Runfola\textsuperscript{1},
Eleonora Bruno\textsuperscript{4,5},
Maria Cristina D'Oca\textsuperscript{1,2},
Massimo Midiri\textsuperscript{4},
Francesca Lizzi\textsuperscript{6}
Ian Postuma\textsuperscript{7},
Marco D'Amelio\textsuperscript{4,5},
Alessandro Lascialfari\textsuperscript{7,8},
Alessandra Retico\textsuperscript{6},
Maurizio Marrale\textsuperscript{1,2,*}\\

\bigskip
\bf{1} Department of Physics and Chemistry Emilio Segré, University of Palermo, Viale delle Scienze, edificio 18, Palermo, 90128, Italy
\\
\bf{2} National Institute for Nuclear Physics, Catania Division, Via Santa Sofia,64, Catania, 95123, Italy
\\
\bf{3} Department of Biological, Chemical and Pharmaceutical Sciences and Technologies, University of Palermo, Viale delle Scienze, edificio 16, Palermo, 90128, Italy
\\
\bf{4} Department of Biomedicine, Neurosciences and Advanced Diagnostics - University of Palermo, Via del Vespro 129, Palermo, 90127, Italy
\\
\bf{5} University-Hospital Paolo Giaccone of Palermo, Via del Vespro 129, Palermo, 90127, Italy
\\
\bf{6} National Institute for Nuclear Physics, Pisa Division, Largo B. Pontecorvo, 3, Pisa, 56127, Italy
\\
\bf{7} National Institute for Nuclear Physics, Pavia Division, Via Agostino Bassi, 6, Pavia, 27100, Italy
\\
\bf{8} Department of Physics, University of Pavia, Via Agostino Bassi, 6, Pavia, 27100, Italy
\\
\bigskip
*Corresponding email: maurizio.marrale@unipa.it
$\dagger$ Co-first authors 

\end{flushleft}

\section*{Abstract}
%%%
In the last years \textit{in-vivo} tractography has assumed an important role in neurosciences, for both research and clinical applications such as non-invasive investigation of brain connectivity and presurgical planning in neurosurgery. In more recent years there has been a growing interest in the applications of diffusion tractography for target identification in functional neurological disorders for an increasingly tailored approach. The growing diffusion of well-established neurosurgical procedures as deep brain stimulation (DBS), radiofrequency ablation (RFA) and stereotactic radiosurgery (STR), or more recently introduced methods as trans-cranial Magnetic Resonance-guided Focused Ultrasound (tcMRgFUS) and MR-guided laser interstitial thermal therapy (MRgLITT), favored this trend. Tractography can indeed provide more accurate, patient-specific, information about the targeted region if compared to stereotactic atlases. On the other hand, this tractography-based approachs is not very physician-friendly, and its heavily time consuming since needs several hours for Magnetic Resonance Imaging (MRI) data processing. 

In this study we propose a novel open-source deep learning framework called DeLTA-BIT (acronym of Deep-learning Local TrActography for BraIn Targeting) for fast target predictions, based on probabilistic tractography. The proposed framework exploits a convolutional neural network (CNN) to predict the location of the Ventral Intermediate Nucleus of the thalamus (VIM). The CNN was trained on the Human Connectome Project (HCP) dataset. The model capability in predicting the VIM location was tested both on the HCP (\textit{internal validation}) and clinical data (\textit{external validation}).
Results from the internal validation have shown good capability in predicting the VIM region (mean $DSC = 0.62\pm 0.15$, mean $sDSC=0.76\pm0.17$) by using just T$_1$ images as input, in a time scale of fraction of second per subject. As for the clinical data, results have been compared with an atlas-based method demonstrating similar performance, but within a significantly shorter timeframe.
\\
The code of the proposed framework is freely available on GitHub.
%%%%
\\
\textbf{Keywords}: Deep Learning; Targeting; transcranial magnetic resonance-guided focused ultrasound surgery; Deep brain stimulation; probabilistic tractography; Thalamic Nuclei; Ventral Thalamic Nuclei;
Nucleus Ventralis Intermedius (VIM)

% now start line numbers
%\linenumbers

% the * after section prevents numbering
\section{Introduction}

The increasing use of non-invasive techniques to study central nervous system connectivity has opened new opportunities in neuroscience. In-vivo tractography via MRI has become essential for research and clinical applications, including brain connectivity studies and presurgical planning in neuro-oncology \cite{le2015diffusion}. Diffusion tractography is also gaining importance in neurofunctional disorder treatments, enabling previously unimaginable levels of precision and allowing clinicians to benefit from imaging-assisted targeting enhancing precision in procedures like deep brain stimulation (DBS) \cite{rissardo2023deep,groppa2023perspectives}, radiofrequency ablation (RFA) \cite{taira2018stereotactic}, stereotactic radiosurgery (STR) \cite{cahan2018radiosurgical}, transcranial MRI-guided focused ultrasound (tcMRgFUS) \cite{peters2023outcomes}, and MR-guided laser interstitial thermal therapy (MRgLITT) \cite{ordaz2023single}.

Essential tremor (ET), the most common movement disorder in adults, significantly impacts daily activities due to involuntary trembling of the hands, head, voice, or limbs, affecting quality of life \cite{louis2010common,chandran2013quality}. A potential cause is abnormal activity in the central tremor network, though its exact pathogenesis remains unclear \cite{brittain2015distinguishing}. Treatment depends on symptom severity and patient needs, ranging from behavioral therapy to medications and surgery \cite{doi:10.1056/NEJMcp1707928}. When medication fails, surgical intervention targeting the ventral intermediate nucleus (VIM) of the thalamus can suppress tremor \cite{fang2016essential}. DBS remains the gold standard \cite{gardner2013history}, but tcMRgFUS has emerged as a non-invasive alternative offering precise, incision-free brain lesioning under MRI guidance \cite{moser2013mr,gallay2018safety} since has achieved increasing recognition at both international and national levels \cite{ferreira2019mds, pouratian2020american, welton2021essential,stieglitz2021consensus,Tremor22}.

Accurate VIM targeting is critical, given its small size ($4 \times 4 \times 6$ mm$^3$) \cite{sammartino2016tractography}. Traditional methods rely on stereotactic atlases or AC-PC-based coordinates, but these approaches do not account for anatomical variability \cite{ghanouni2015transcranial}. Advanced MRI techniques, including susceptibility-weighted imaging (SWI) and diffusion tractography, have been explored for patient-specific targeting \cite{shah2020advanced}. Among these, probabilistic tractography is the most promising, as it models white matter pathways while incorporating uncertainty in MRI diffusion data \cite{basser1994mr,behrens2003characterization, lehman2020mri}. Studies have shown its effectiveness in VIM targeting, correlating well with clinical outcomes \cite{krishna2019prospective,feltrin2022focused,tian2018diffusion}.

Despite its advantages, probabilistic tractography is computationally demanding, requiring hours per patient, and is not user-friendly. Machine learning (ML) and deep learning (DL) offer data-driven solutions to streamline tractography \cite{litjens2017survey}. Recent DL models have successfully estimated fiber orientations and generated tractography maps \cite{neher2015machine,benou2019deeptract}. Various architectures, including CNNs and Transformers, have been developed to improve fiber tracking \cite{wasserthal2018tractseg,hosseini2022cttrack}.

In this study, we introduce DeLTA-BIT (Deep-learning Local TrActography for BraIn Targeting), an open-source DL framework designed for fast and accurate tractography-based ROI identification. We present the results of training a 3D UNnet to predict VIM ROI using only T$_1$ images.

The UNnet was first trained on the dataset of the Human Connectome Project (HCP). Two validation steps have been carried out: i) \textit{internal validation}, where it was tested the model capability in predicting ROIs obtained by probabilistic tractography on the HCP data; ii) \textit{external validation}, where it was tested the model capability in predicting the VIM location on clinical data, along with a comparison with an atlas-based method (THOMAS, \cite{su2019thalamus}).
VIM predictions are achieved in a very short time (fraction of second per subject, to be compared with tens of minutes with THOMAS), making the proposed method promising for live targeting in MR-guided procedures.  

\subsection{Data source}
\label{data_source_section}

Data from the Human Connectome Project (HCP) (\cite{van2013wu,sotiropoulos2013effects,milchenko2013obscuring,glasser2013minimal}), relative to approximately 1200 young adult subjects, have been used for this study. Only subjects whose T$_1$ and DTI images were both present and acquired using the same scanner specifics (3 Tesla scanner) have been selected. Thus, the final size of the dataset is 1064 subjects, of which 800 were included in the training set and 264 in the test set , the latter for internal validation.
In order to identify the VIM, within the left thalamus, the same tractography pipeline has been applied to all subjects images.

Clinical data from 7 ET patients who underwent unilateral MR-guided focused ultrasound thalamotomy (targeting the presumed Vim) for tremor control were used for the external validation analysis. Clinical outcome assessments were performed by a neurologist with $\geq 20$ years of experience in movement disorders. Only patients achieving an optimal clinical outcome (defined as a $\geq 80\%$ reduction in the treated-side at The Essential Tremor Rating Assessment Scale (TETRAS) score) assessed at least one year after treatment were included, under the assumption that such sustained, marked improvement reflects accurate targeting and effective ablation of the VIM nucleus. Patients included in these analyses were retrieved from a previously conducted study at the University-Hospital ``Paolo Giaccone'' of Palermo funded by the Italian Ministry of Health (``Ricerca Finalizzata 2016'', grant no. GR-2016-02364526; principal investigator: Dr.\ Cesare Gagliardo). The study protocol received approval from the local Institutional Review Board (approval code N.\ 04/18). All participants provided written informed consent prior to enrollment, and all procedures were conducted in accordance with the principles of the Declaration of Helsinki. In particular axial 3D T1-weighted Inversion Recovery (IR) BRAin Volume (BRAVO) pulse sequence acquired using a 1.5T scanner (GE Signa HDxt with an 8-channel head coil) with a 1mm isotropic resolution were used (scan parameters: slice thickness 1mm; repetition time 12.4ms; echo time 5.2ms; inversion time 1866ms; matrix 256x256; number of excitation 1; filed of view 25.6x25.6). Each subject underwent both a pre-treatment screening MRI, routinely used for treatment planning, and a 48-hour post-treatment follow-up scan. The screening and follow-up images were subsequently rigidly co-registered, and binary lesion masks were segmented from the follow-up scans.

Lesion segmentation was manually performed on 48-hour postoperative 3D T1 IR BRAVO pulse sequences. Segmentations strictly included the well-defined T1 hypointense core corresponding to the previously characterized Zone II (cytotoxic edema), while explicitly excluding the surrounding Zone III (vasogenic edema) (\cite{wintermark2014imaging,gagliardo2020intraoperative}).
All segmentations were generated in ITK-SNAP version 4.2.2 \cite{yushkevich2006user} using the “pre-segmentation thresholding” tool by first defining upper and lower intensity thresholds, after which a seed bubble was placed within the lesion to initialize the differential-equation–based contour-evolution algorithm. All delineations were performed by a neuroradiologist with $>$12 years of experience in neuroanatomical segmentation and $>$10 years of experience as primary operator in trans-cranial MRgFUS procedures.

Correlation between post-treatment clinical outcomes and the performance of the evaluated algorithms was assessed using the TETRAS score, considering only the hemi-score corresponding to the treated side.

External validation was performed by using the screening images as input for both the NN and the THOMAS pipelines, while the manually generated lesion masks served as the ground truth (GT).

\subsection{THOMAS Atlas}

Thalamus-Optimized Multi-Atlas Segmentation (THOMAS) is a thalamus specific atlas-based pipeline able to identify 12 nuclei and sub-nuclei (22 ROIs plus thalamus ROI for each hemisphere), using just T1 images, as described in \cite{su2019thalamus}. It was run in a docker container with operating time of 15-30 min per patient on our workstation (DELL Precision Tower, 128 GB RAM, Intel® Xeon(R) Platinum 8268 CPU @ 2.90GHz × 48).   
We used the THOMAS-derived left thalamus segmentation to perform probabilistic tractography, and adopted the VLPv ROI (i.e., the inferior portion of the ventral lateral posterior nucleus, as defined by Su et al., 2019 \cite{su2019thalamus}) as a surrogate for the VIM in the clinical dataset.

\subsection{Standard pipeline for probabilistic tractography}
The probabilistic tractography estimation requires multiple steps performed through the \textit{FreeSurfer} and \textit{FSL} tools (\cite{fischl2012FreeSurfer,smith2004advances,woolrich2009bayesian,jenkinson2012fsl}) and python code suitably developed using the numpy, nibabel and scikit-learn libraries \cite{harris2020array,pedregosa2011scikit}.

\begin{figure}[!h]
    \centering
    \includegraphics[width=14cm]{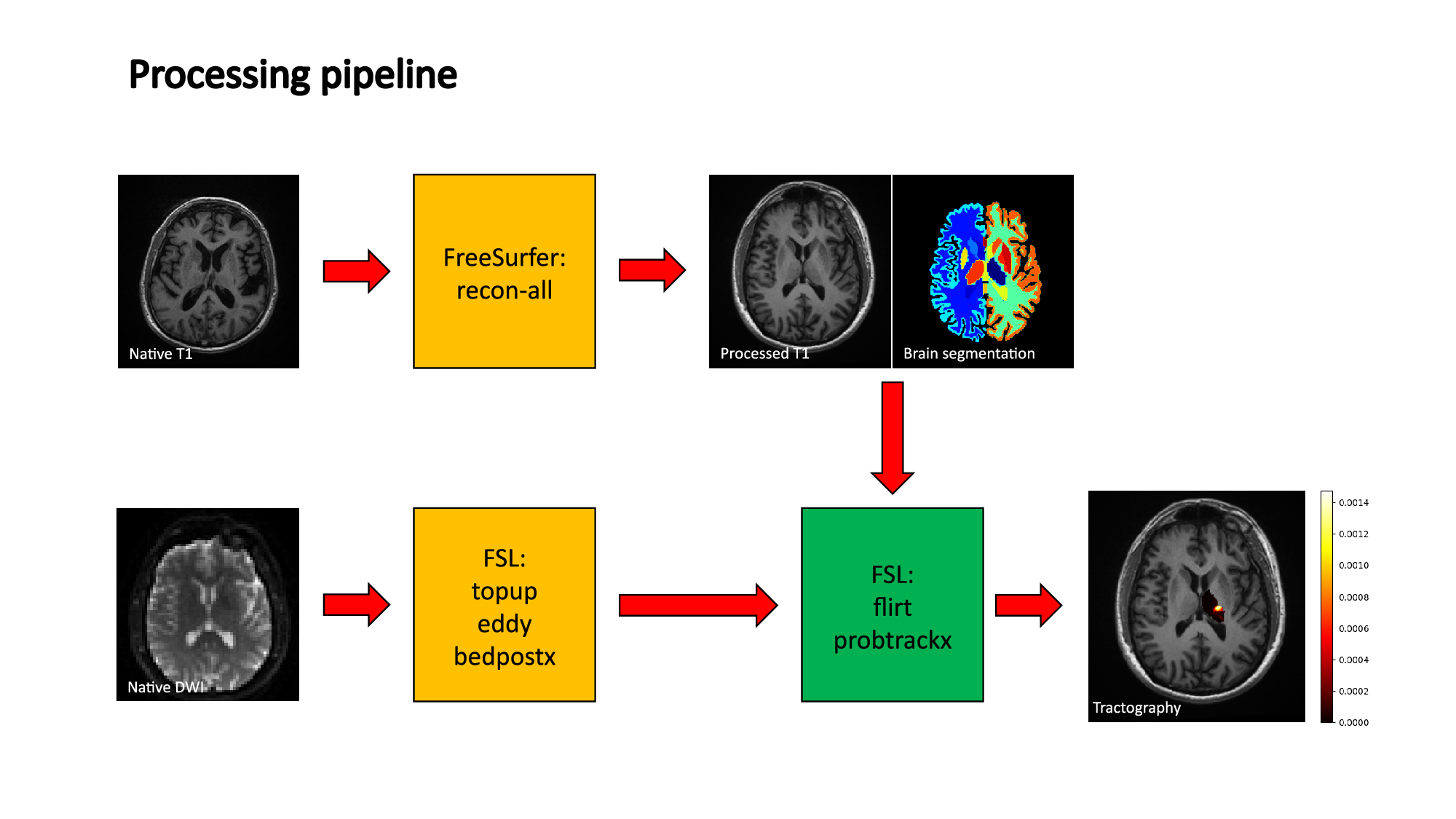}
    \caption{Schematic representation of the probabilistic tractography pipeline.}
    \label{fig:pipeline}
\end{figure}

First, MRI data must be converted in NIfTI format. Subsequently, T\textsubscript{1} and DTI images are processed separately. Results coming from both processing steps are then combined to run the PROBTRACKX routine, as shown in Figure \ref{fig:pipeline}.

\subsubsection{T$_1$ preprocessing}
T1 images are essential for generating binary masks of brain regions requested for probabilistic tractography. Prior to brain region extraction, preprocessing can be conducted using the \textit{Freesurfer} software. The complete pipeline, executed via the \texttt{recon-all} command, involves noise and motion correction, intensity normalization, and labeling of both cortical and subcortical brain regions. Subsequently, binary masks for both cortical and subcortical regions can be derived using additional FreeSurfer commands: \texttt{mri\_annotation2label}, \texttt{mri\_label2vol}, and \texttt{mri\_binarize}, this last utilizing the segmentation files produced by \texttt{recon-all} to create binary masks. Lastly, given FreeSurfer's unique data representation, a data reorientation to MRI standard can be performed using the FSL command \texttt{fslreorient2std}.
The SUIT atlas \cite{diedrichsen2006spatially, diedrichsen2009probabilistic, diedrichsen2011imaging, diedrichsen2015surface} has been exploited for the right dentate nuclei segmentation.

\subsubsection{DTI preprocessing}
Since DTI images are prone to noise and artifacts, DTI preprocessing is mandatory. \textit{FSL} tools (\texttt{topup}, \texttt{eddy})  are employed for motion and eddy current corrections, as well as noise reduction. After these first steps, the
Orientation Fiber Distribution (OFD) at each voxel was estimated using \texttt{BEDPOSTX} (Bayesian Estimation of Diffusion Parameters Obtained using Sampling Techniques, where X stands for modelling Crossing Fibres, \cite{behrens2003characterization}). 

\subsubsection{Probabilistic tractography and connectivity map}
\label{prob_tract}
\begin{figure}[!h]
\centering
\begin{tabular}{c}
 \includegraphics[width=13.5cm]{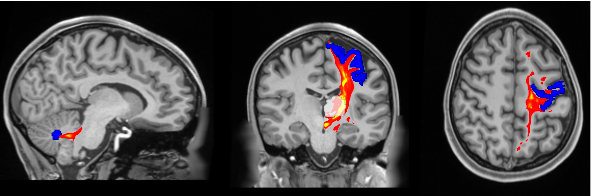} \\ (a) \\ \includegraphics[width=10cm]{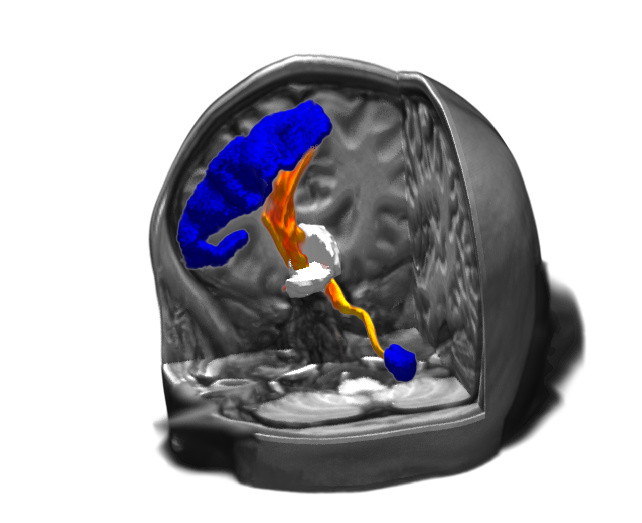} \\ (b) \\
\end{tabular}
    \caption{Probabilistic tractography streamlines for the DRTC tract. In blue the left-M1-cortex and the right dentate nucleus ROIs. In white the left thalamus. In red-to-yellow scale the DRTC tractogram obtained by merging the dentate-thalamus-M1 tractography in both directions. Panel a, from left to right: sagital, coronal and axial projections. Panel b: rendering of the 3D image. }
    \label{fig:streamline}   
\label{fig:streamlines}
\end{figure}

\begin{figure}[!h]
   \centering
  \includegraphics[width=10cm]{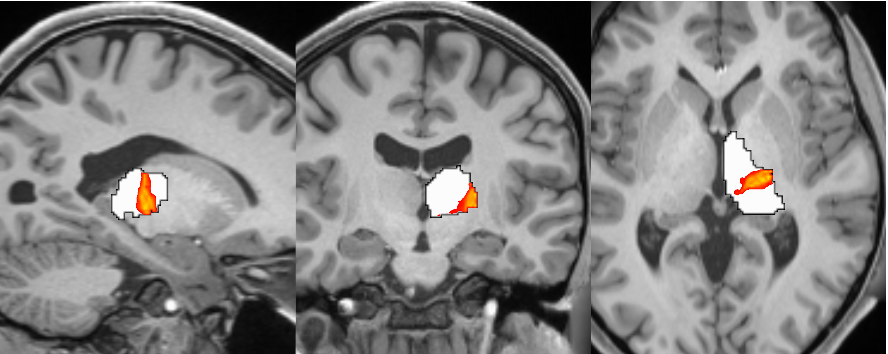}
 \caption{M1 connectivity map obtained setting the left thalamus as seed and the M1 as target region. As is well known, the VIM is located within this area.}
    \label{fig:M1_connectivity}
\end{figure}

Probabilistic tractography has been performed using \texttt{probtrackx2\_GPU} (the GPU version of \texttt{probtracx}, \cite{behrens2007probabilistic,hernandez2019using}). This tool produces sample streamlines, drawing an orientation from the voxel-wise BEDPOSTX distributions and taking a step in this direction. In other words, these streamlines follow a random walk exploiting the information of the water diffusion tensor for each voxel. \\
The process starts from a seed region and ends when a termination criterion is reached. A termination criterion consists in reaching a target area, e.g. a cortical region, or arresting the process when the streamline ends in an avoided region, e.g. brain ventricles.  Additionally, a region can be selected as a way-point, filtering out unuseful streamlines.\\
For each subject, three different tractographies were obtained, by varying seed and target regions: i) left thalamus-to-left M1 cortex; ii) right dentate nucleus-to-left M1 cortex including the left thalamus as a way-point; iii) left M1 cortex-to-right dentate nucleus including the left thalamus as a way-point. The last ones were used to identify the Dentato-Rubro-Thalamo-Cortical (DRTC) track \cite{akram2018connectivity}, see figure \ref{fig:streamlines}.
Furthermore, keeping track of the several seed-to-target paths and counting how often a seed voxel is included in a streamline, a connectivity map can be constructed. In our case, left thalamus-to-left M1 cortex tractography was exploited to evaluate the correspondent connectivity map, shown in Figure \ref{fig:M1_connectivity}.

\subsection{VIM identification}
\label{VIM_def}
\begin{figure}[h!]
    \centering
    \includegraphics[width=10cm]{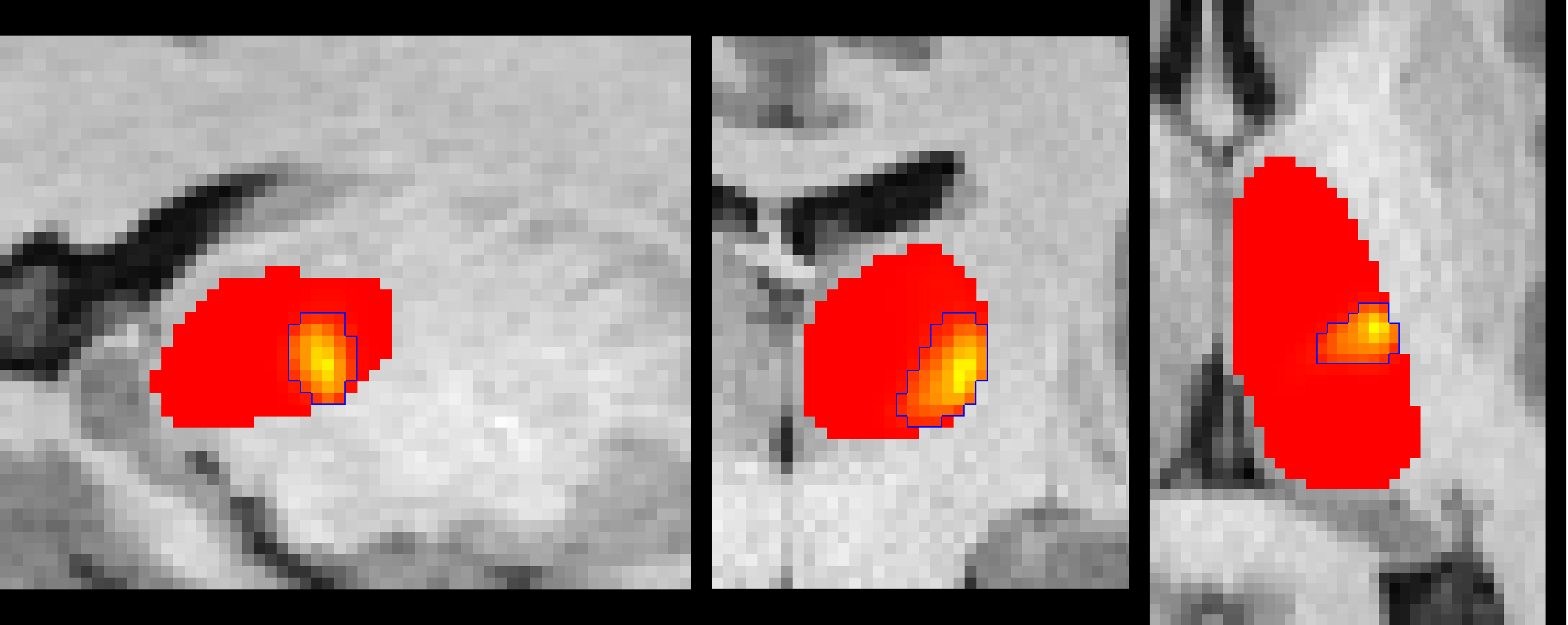}    
    \caption{Identification of the VIM region. The VIM ROI was obtained by applying a mean filter (kernel size $ = 5\times5\times5$) to the raw distribution obtained by intersection. The contour of the obtained binary mask is highlighted in blue.}
    \label{fig:VIM_loc}
\end{figure}

It is known that the VIM receives input from the contralateral dentate nucleus and projects to cortical areas of the ipsilateral motor network, primarily the primary motor cortex (M1) \cite{gallay2008human,mollink2016dentatorubrothalamic, calabrese2015postmortem, yang2022tractography}. For this reason, the VIM location was identified by intersecting the thalamus region, the M1 connectivity map, and the tractogram obtained by merging the dentate-thalamo-M1 tractographies in both directions (see section \ref{prob_tract}).
Since probabilistic tractography is affected by statistical noise, a mean spatial filter was applied to the resulting distribution. In order to obtain a binary mask of the VIM location, a threshold (10\% of the maximum intensity) was applied. Figure \ref{fig:VIM_loc} shows the obtained distribution and the contour of the VIM.

\subsection{AI model}
In what follows, we provide details on the AI model we developed, including the NN architecture, the data structure, and the training protocol.

\subsubsection{Neural Network}
\begin{figure}[ht!] 
\centering
    \includegraphics[scale=.4]{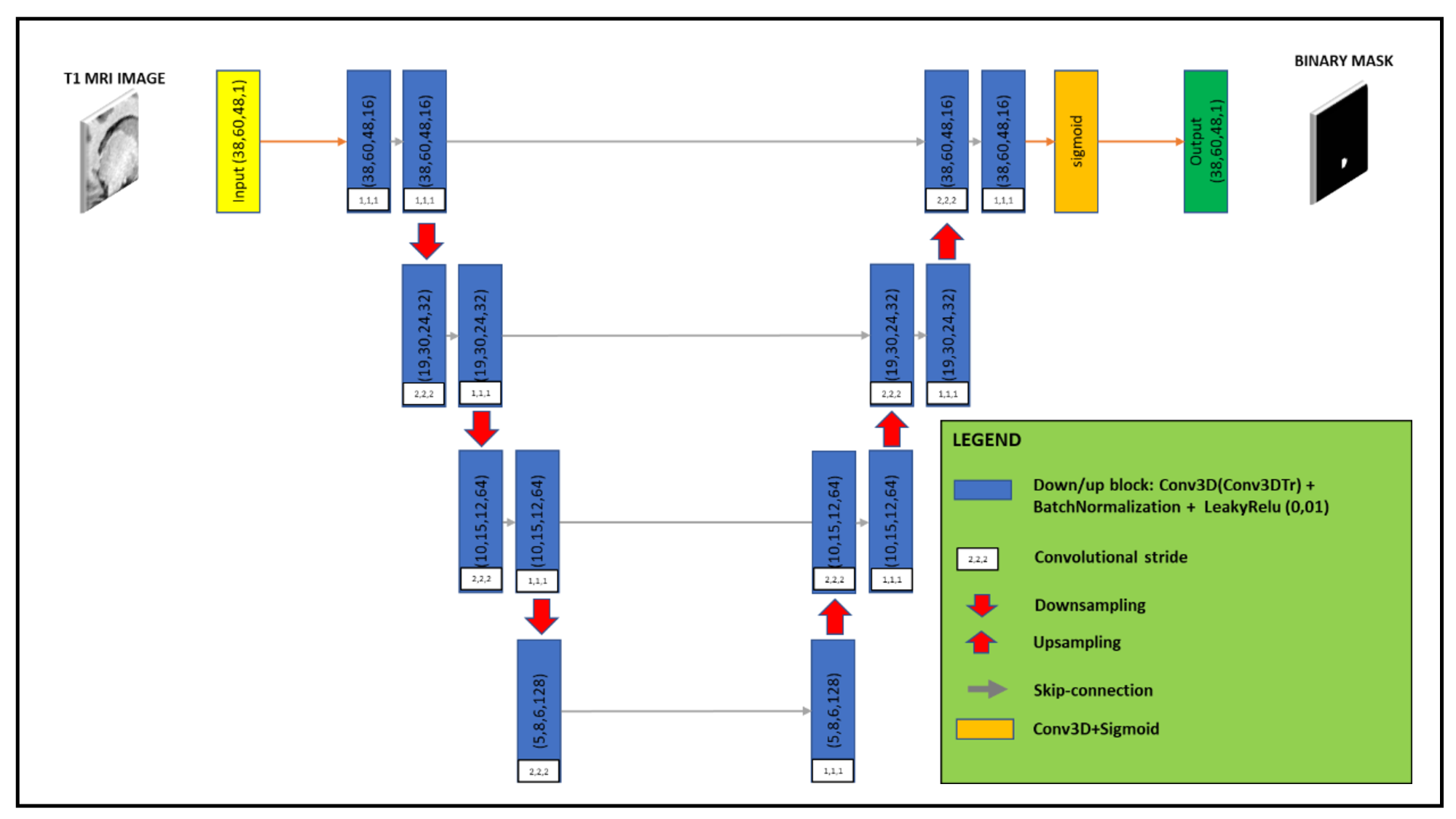}
\caption{Schematic representation of the U-Net neural network. This network exhibits a symmetric shape, working as an encoder-decoder: the left side corresponds to the encoder, while the right side represents the decoder. The network is organized into levels, with each level containing two convolutional blocks. Each convolutional block consists of a 3D convolutional layer (kernel size 3), followed by a LeakyReLU activation and Batch Normalization layers. The numerical values within the blocks (in parentheses) denote the block outputs shape. Specifically, the first three numbers indicate spatial dimensions, while the last number represents the number of feature maps. Additionally, the numbers at the bottom of the blocks correspond to the convolutional stride. A stride of 1 maintains the same spatial size for the layer output, while a stride of 2 spatially reduces the output by half. At each level, the initial convolutional block employs a stride of 2 (except for the first level), effectively halving the input size in all directions. This reduction in spatial dimensions is compensated by doubling the number of feature maps. The output layer consists of a convolutional block with a sigmoid activation function, yielding output values within the range 0-1, representing the probability for a voxel to belong to the VIM region.}
\label{network}
\end{figure}

Our network's architecture employs a 3D U-Net, a fully convolutional neural network (CNN), as depicted in Figure \ref{network}. The U-Net, renowned for its effectiveness in segmentation tasks, has also found utility in regression tasks and Generative Adversarial Network (GAN) models for image translation (\cite{isensee2021nnu,isola2017image}). Notably, the U-Net operates as an 'encoder-decoder' model, incorporating skip connections to mitigate vanishing gradients. 

The network input traverses through the architecture to produce an output of the same size. The network architecture consists of 4 levels, with two convolutional blocks per level. Each convolutional block comprises a 3D convolutional layer (kernel size: $3$), a LeakyReLU activation layer (with $\alpha=0.01$), and Batch Normalization layer. In the encoding phase (left side of the network in Figure \ref{network}), convolutional blocks progressively reduce the spatial dimensions of the layer inputs while introducing new feature maps. At each level, the input spatial size is typically halved, while the number of feature maps doubles. This process continues until the fourth level, where the input becomes compact (size: $5\times8\times5$ voxels) but it has a large number ($128$) of feature maps. This level is also said \textit{bottle neck} and its representation is said \textit{latent space}. Here, the network has completed encoding and transits to decoding (right side of the network in Figure \ref{network}). During decoding, the network reverses the process: it enlarges the layer input spatially and reduces the feature maps using 3D transposed convolutional layers. The output layer consists of a convolutional block with a sigmoid activation function, yielding output values within the range 0-1, representing the probability for a voxel to belong to the region being segmented. \\
The network expects a 3D image as input — T1-normalized MR image in our case — and attempts to predict the spatial localization of the VIM, as defined in Section \ref{VIM_def}.

\begin{figure}[!h]%\centering
\includegraphics[scale=.35]{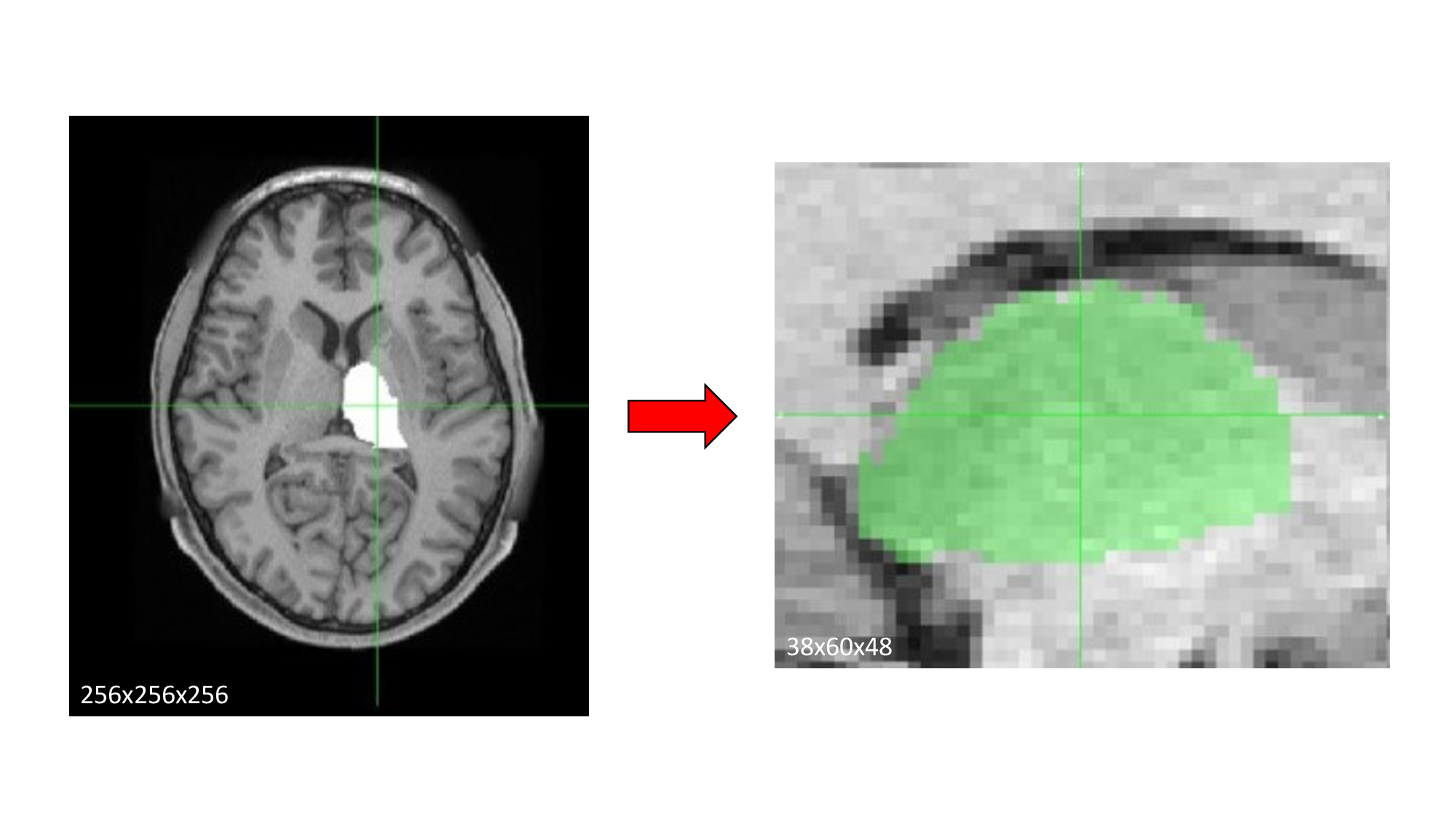}
\caption{Bounding box for left thalamus, suitable for all dataset subjects (total volume of $\mathbf{38\times60\times48}$ voxels).}
\label{bbox}
\end{figure}

Prior to training the CNN, a comprehensive analysis of computational complexity was conducted. A statistical investigation into the physical characteristics of the thalamus, including volume and surface area, revealed that the mean volume of the thalamus in a hemisphere is approximately $8600\ mm^3$, equivalent to roughly $0.05 \%$ of a standard MRI image (typically sized at $256\times256\times256$ mm$^3$). To address this, a bounding box was deliberately selected. Given that all Human Connectome Project (HCP) images were aligned to the MNI152 standard, they share a consistent frame of reference. The voxel coordinates for the bounding box were determined by randomly selecting one hundred subjects and loading their thalamus binary masks. Subsequently, the smallest box that encapsulated the largest thalamus across all selected subjects was chosen. A bounding box measuring $38\times60\times48$ voxel (as depicted in Figure \ref{bbox}) was identified. Notably, this bounding box is significantly smaller than a standard MRI image, making it well-suited for CNN training purposes. 

\subsection{Model training}

The NN comprises a total of $1,462,113$ parameters, of which $1,460,705$ are trainable and initialized randomly using a normal distribution. We employed the \textit{Adam algorithm} as the optimizer and the \textit{combo loss} as the loss function, which combines cross-entropy and dice loss.

Given an ample number of subjects available for training, we performed a single validation split to monitor the training process for each model. Consequently, the training set consisted of $704$ subjects, while the validation set comprised $96$ subjects (randomly selected). Due to the large size of the training set and memory constraints, we employed a data generator using a batch size of $32$ subjects. The data generator also facilitated data normalization, online data augmentation, and random shuffling at the end of each epoch. Data normalization ensures that inputs fall within the range of $0$ to $1$, a common practice in deep learning to facilitate training. Data augmentation mitigates overfitting and provides additional data when the training set is limited. Our augmentation techniques included small-angle rotations, translations, flips, scaling, and the addition of random Gaussian noise. During each epoch, the network encountered different data due to these random transformations. \\
To prevent the network from consistently seeing the same sequence of data (which could lead to overfitting), we applied random shuffling to the data list at the end of each epoch. Additionally, we employed two regularization techniques: i) saving a model checkpoint during training every time the model reaches a new highest score on the validation set and ii) applying the early stopping algorithm, with 200 patience epochs.\\
Our hardware setup consisted of a Dell workstation with Intel Xeon(R) Platinum $8268$ CPU $@$ $2.90$GHz, $48$ cores, $128 GB$ of RAM, \texttt{NVIDIA RTX A5000} GPU with $24 GB$ of memory. We initially set $1000$ epochs, but the training was stopped at 284 epochs (by the Early stop regularization criterion), taking approximately $2$ hours.
The code of this model, both for training and testing, is freely available on Github at the link \url{https://github.com/mromeo1992/delta-BIT}.

\subsection{Evaluation methods}
\label{evaluation}
The prediction ability of the model has been evaluated on the test sets (HCP test set and clinical data). The metric we used is the Dice Similarity  Coefficient (DSC), defined as:
\begin{equation}
    DSC=2\frac{| X \cap Y |}{|X|+|Y|}
    \label{DSC}
\end{equation}
where $X\ and\ Y$ are two ROIs to compare and $|X|$ represent the size of the ROI $X$.\\
Additionally, since the VIM region is very small compared to the size of the image, we included two more metrics in model evaluation: i) the surface Dice Similarity Coefficient (sDSC) and ii) the distance between the centers of the mass of the true and predicted VIM (D$_{COM}$).
The sDSC has a similar definition to the DSC, or \textit{Volume} DSC, except that only the voxels along the surface are considered. This requires a tolerance value to detect voxels located along the surface of a volume, which we set to 1 mm.

Finally, in the external validation, two additional metrics were considered. The first is the True Positive Rate (TPR), also referred to as $Recall$. Given a predicted ROI $X$ and a GT ROI $Y$, we define $TP$ (\textit{True Positives}) as the voxels classified as positive (belonging to the ROI) in both $X$ and $Y$, $FP$ (\textit{False Positives}) as the voxels classified as positive in $X$ but not in $Y$, and $FN$ (\textit{False Negative}) as the amount of positive voxels in $Y$ missed in $X$. Based on these definitions, the TPR is expressed as:
\begin{equation}
\mathrm{TPR} = \frac{TP}{TP + FN}.
\end{equation}
In other words, Recall can be seen as the portion of GT volume recovered in the prediction.\\
In similar way, the \textit{Precision} of a segmented region is evaluated as:
\begin{equation}
    Precision=\frac{TP}{TP+FP}.
\end{equation}
Precision corresponds to the proportion of the predicted volume that falls inside the GT volume. 
It is important to evaluate these two metrics jointly, since a predicted region whose volume is much larger than the GT and fully contains it would yield a TPR equal to 1, but a very low Precision (given by the ratio GT volume/predicted volume).
Conversely, a region whose volume is entirely contained within the GT would result in high Precision but low TPR (given by the ratio predicted volume/GT volume).
For the sake of completeness, using the same notation DSC can be rewritten as
\begin{equation}
    DSC=2\frac{TP}{FP+FN+2\ TP}.
\end{equation}

\section{Results}
 \subsection{Internal validation}
\begin{figure}[h!]
    \centering
    \begin{tabular}{cc}
    Subject (a) & \includegraphics[width=10cm]{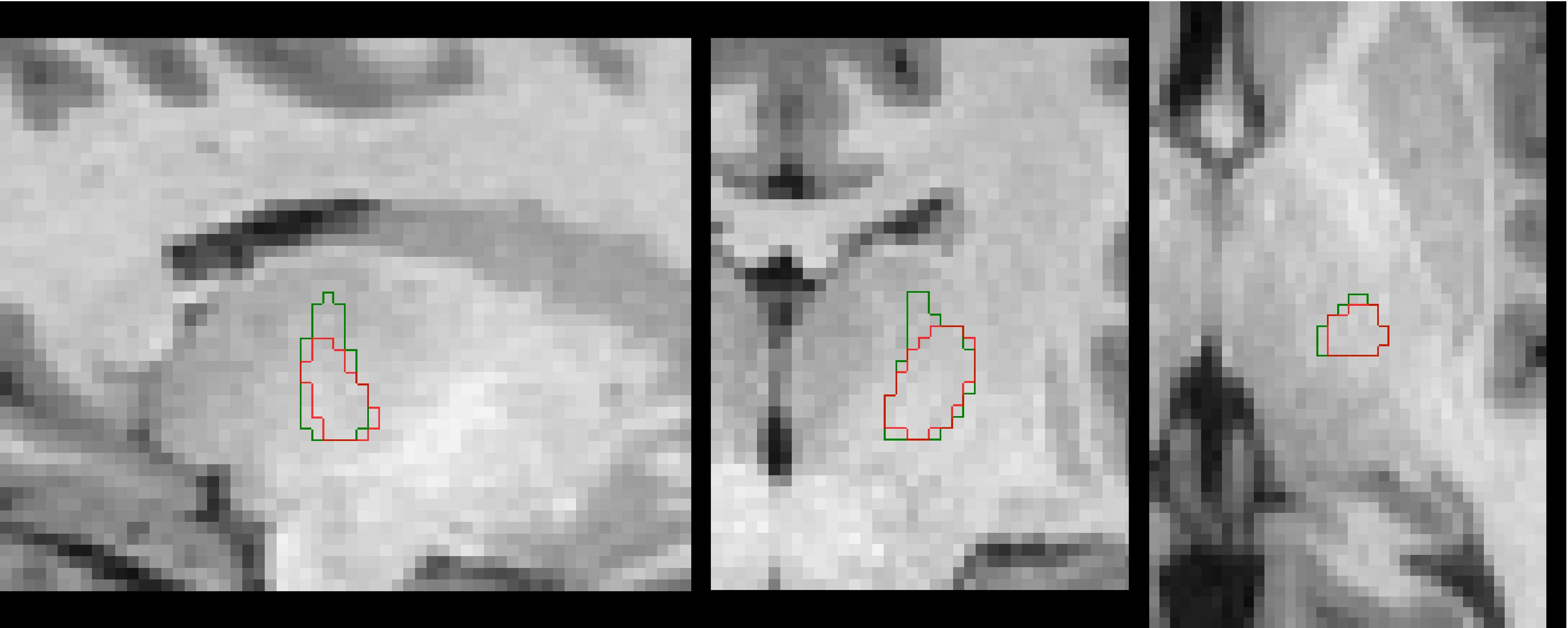} \\
    Subject (b) & \includegraphics[width=10cm]{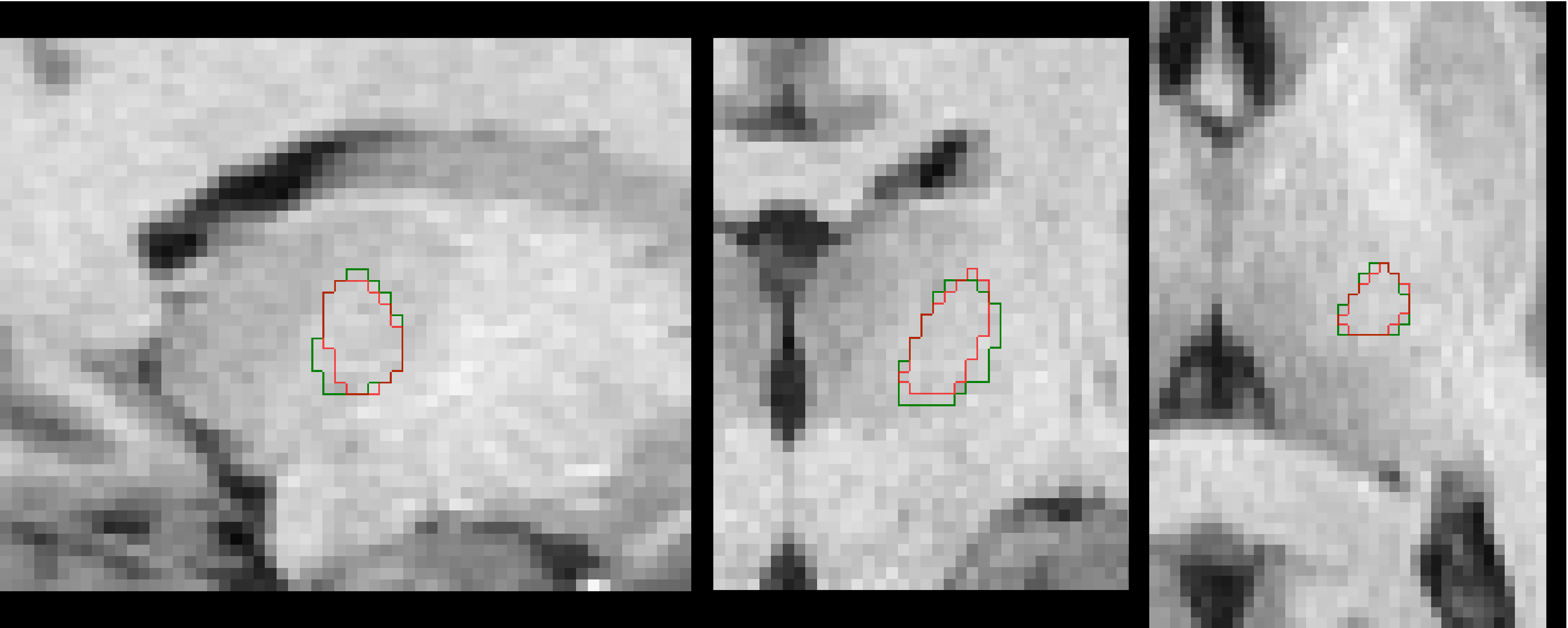} \\
    Subject (c) & \includegraphics[width=10cm]{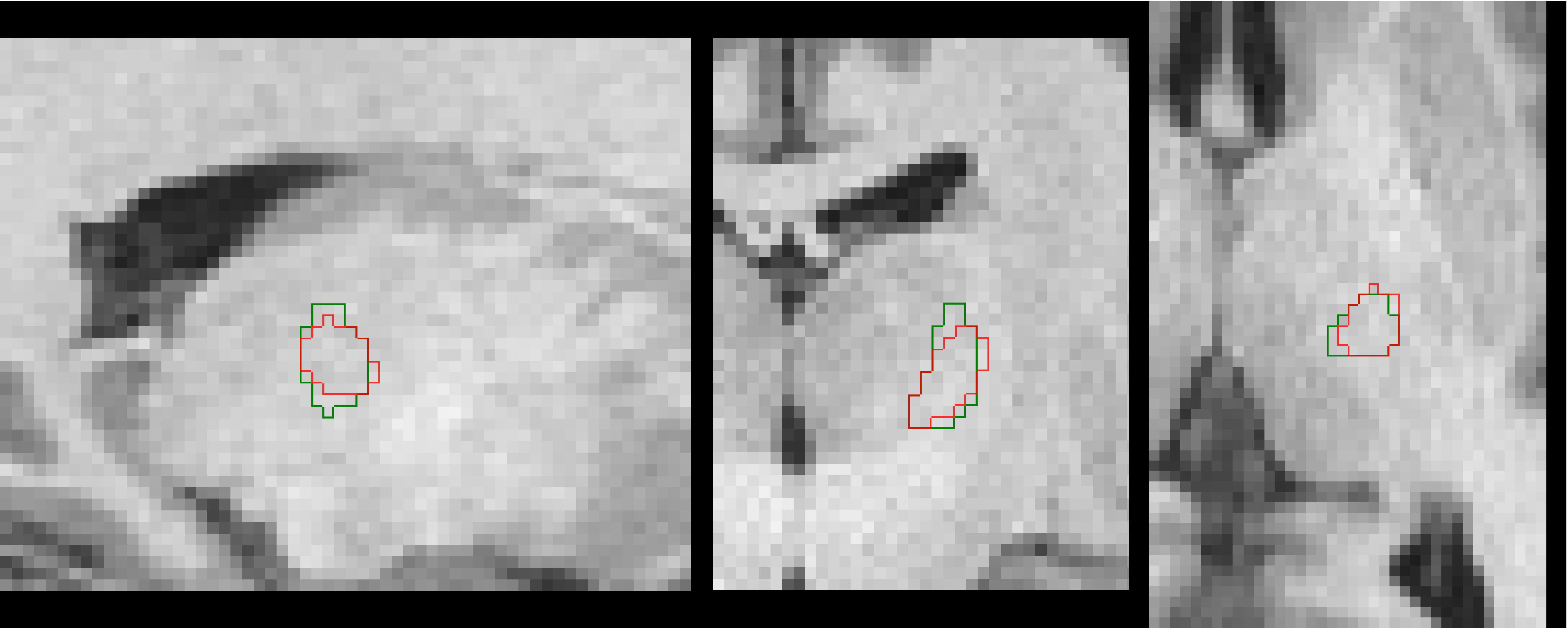} \\
          
    \end{tabular}
    
    \caption{Comparison of VIM ROIs for three representative subjects. From left to right: sagittal, coronal and axial projection. Green: predicted ROI; red: reference ROI.  Subject (a): $DSC = 0.78$, $sDSC = 0.98$;  subject (b): $DSC = 0.80$, $sDSC = 0.99$; subject (c): $DSC = 0.88$, $sDSC = 0.99$.}
    \label{fig:qualitative_overlap}
\end{figure}

In this section, results on the test set (264 samples) are presented. Figure \ref{fig:qualitative_overlap} depicts the overlap between the reference ROI (red contoured region) and the prediction (green contoured region) for three representative subjects, in the sagittal, coronal and axial projections. Although the figure shows only one slice per projection, a good agreement between the GT and the model output is observed.

Table \ref{tab_result1} summarizes the results of the evaluation analysis with different metrics utilized as described in Section \ref{evaluation}, performed over the whole test set. We obtained a mean DSC of $0.62 \pm 0.15$, which is a good result considering the difficulty of localizing such a small area (mean volume of $396\ mm^3$). To perform a more comprehensive evaluation, we also considered the sDSC. This metric assesses the overlap between the surfaces of two ROIs, considering a thickness of $1 mm$. Thus, it allows for evaluating the agreement between the GT and the prediction along the border. For this metric, we obtained a mean value of $0.76\pm0.17$.

\begin{table}[h!]
    \centering
    
    \begin{tabular}{|c|c|c|c|}
        \hline
            & DSC & sDSC & D$_{COM}[mm]$\\
        \hline
         mean & 0.62 & 0.76 & 1.74\\
         \hline
         std & 0.15 & 0.17 & 0.83\\
         \hline
    \end{tabular}
    \caption{Evaluation analysis with different metrics utilized.}
    \label{tab_result1}
\end{table}

\begin{figure}[h!]
    \centering
    \begin{tabular}{c}
    \includegraphics[width=10cm]{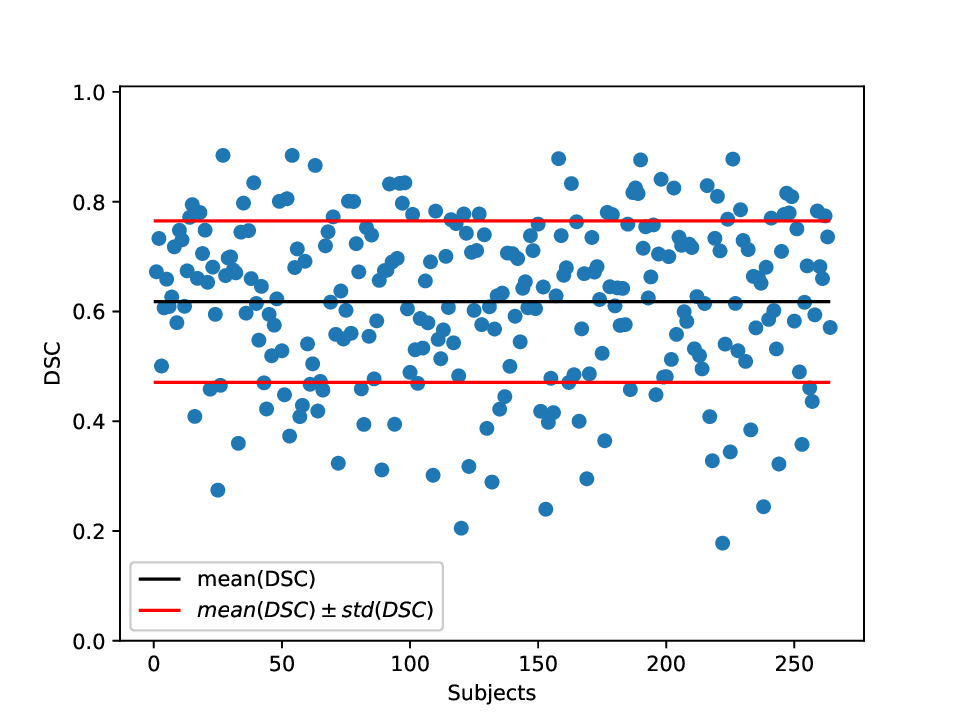} \\ (a)\\ 
    \includegraphics[width=10cm]{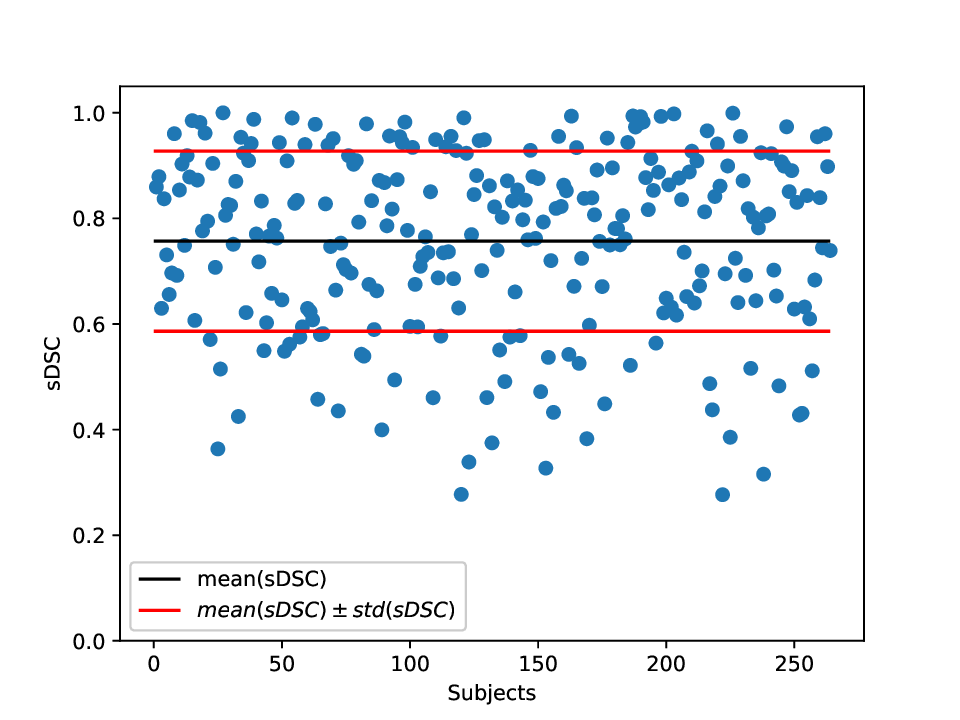}  \\(b)\\

   \end{tabular}
        \caption{Volume (upper panel) and surface (lower panel) Dice similarity coefficient between the predicted and the reference VIM binary masks. Average scores:  $DSC=0.62 \pm 0.15$ and $sDSC=0.76 \pm 0.17$.}
        \label{fig:DSC_sDSC}
\end{figure}

In Figure \ref{fig:DSC_sDSC}a and b we show the dispersion of the results around the mean values.

\begin{figure}[h!]
    \centering
    \includegraphics[width=6.5cm]{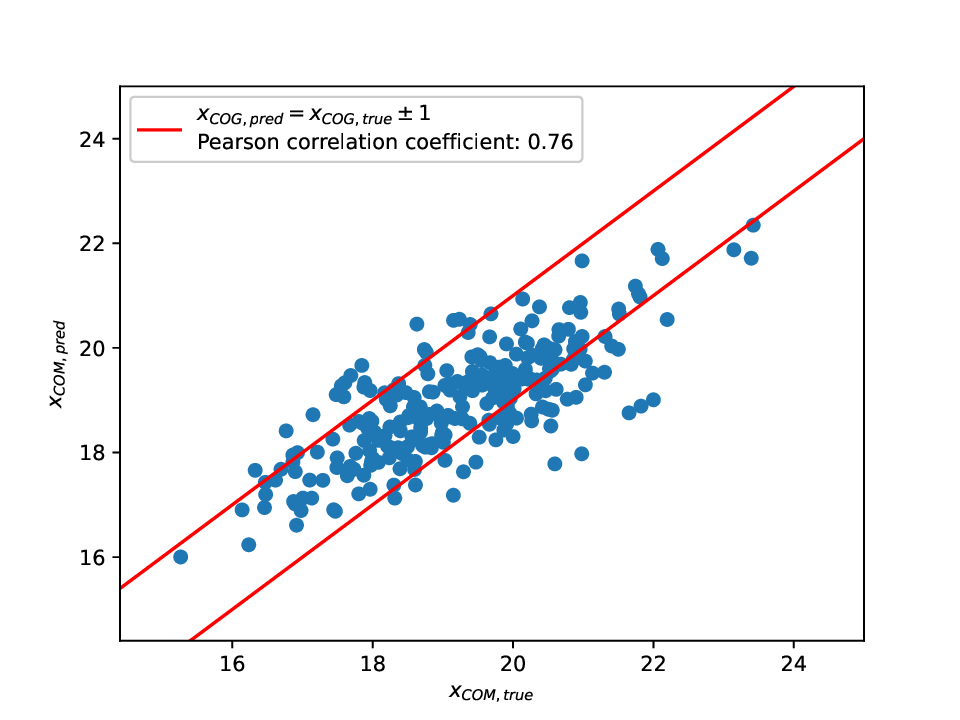}
    \includegraphics[width=6.5cm]{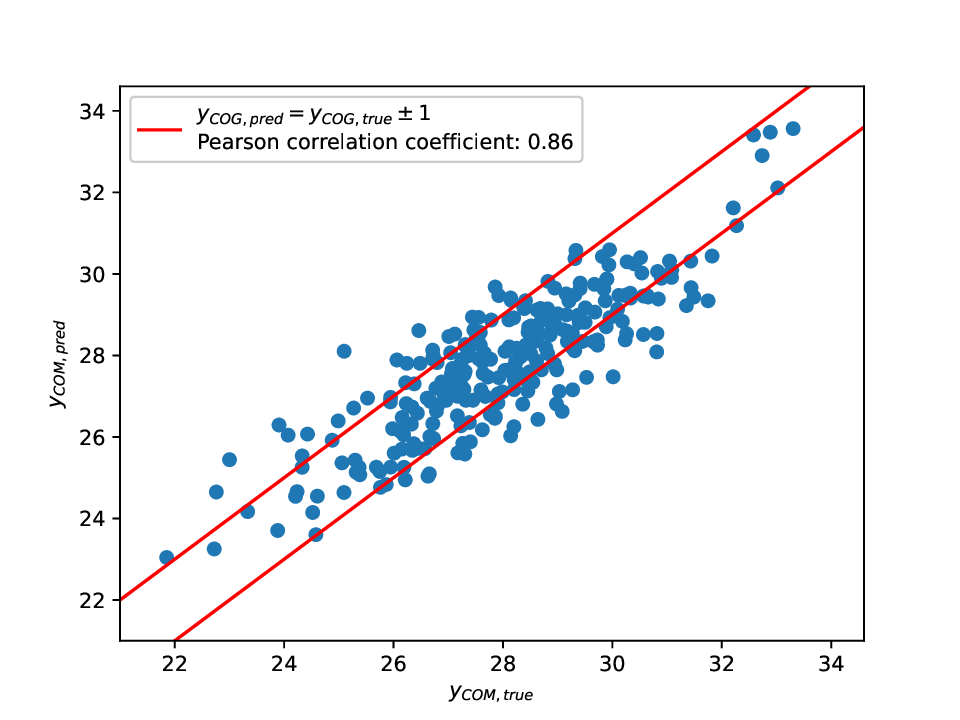}
    \includegraphics[width=6.5cm]{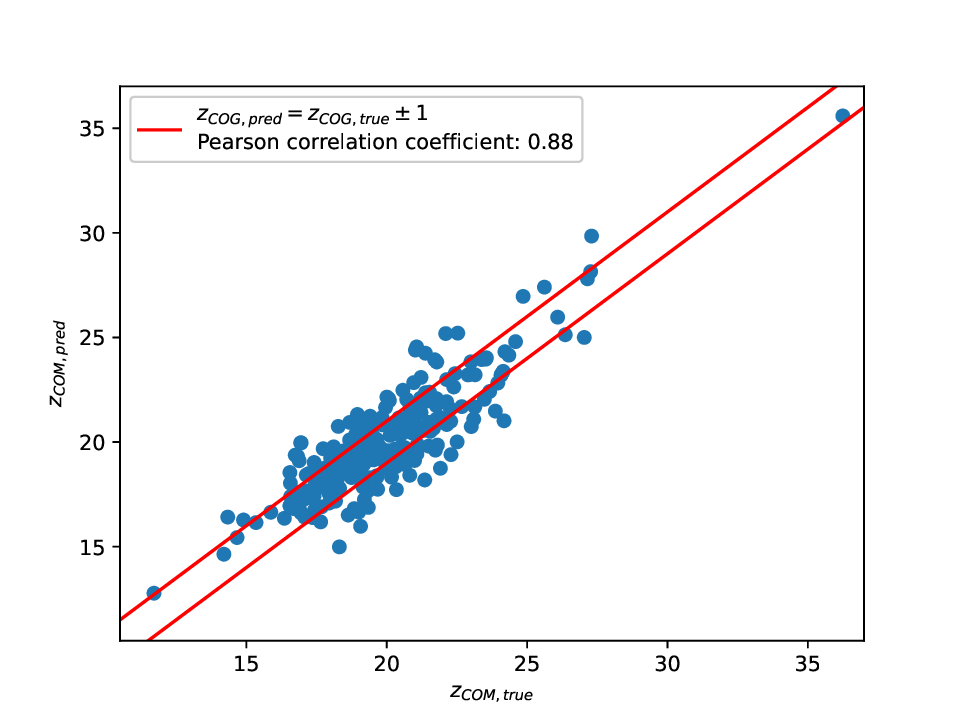}
    \caption{Correlation plot of the coordinates of the centre of mass for the true and the predicted VIM ROIs.}
    \label{fig_COM}
\end{figure}

Finally, as the last evaluation test, we considered the distance between the centers of mass (D$_{COM}$) of the reference and predicted ROIs. We obtained a mean distance of $1.74 \pm 0.83\ mm$, which is comparable to the size of voxels along a direction. The correlation plots for each direction are shown in Figure \ref{fig_COM}.

Besides the model's performance, it is important to evaluate the gain in terms of time required. The standard pipeline requires approximately 2.5 hours for T$_1$ pre-processing and between 2 minutes and 4 hours for DTI pre-processing, depending on the number of gradient acquisitions. Our model requires only a T$_1$ MR image to predict the VIM location, eliminating the need for DTI acquisition and processing. The time required to predict the VIM ROIs for the whole test set (264 subjects) was 5.9 seconds on our hardware. This result is of great importance, as it enables VIM localization during surgical procedures—something not feasible with the standard MR pulse sequences—and the remarkably short computation times make it particularly promising for clinical translation, where lengthy processing often drives the choice toward alternative targeting strategies.

\clearpage

\subsection{External validation}
To evaluate the generalization capability of our AI model, we used clinical data from 7 ET patients who underwent FUS treatment. T1 screening images have been used as input for both the NN and the THOMAS pipeline. T1 follow-up images, where the edema induced by thermal ablation appears as a hypo-intense region, were used to derive reference ROIs (see section \ref{data_source_section}). These ROIs can be seen as GT for the screening images after appropriate co-registration, as they are supported by positive clinical outcomes. Scores obtained by NN and THOMAS are reported and compared in this section.

\begin{figure}[h!]
    \centering
    \includegraphics[width=10.0cm]{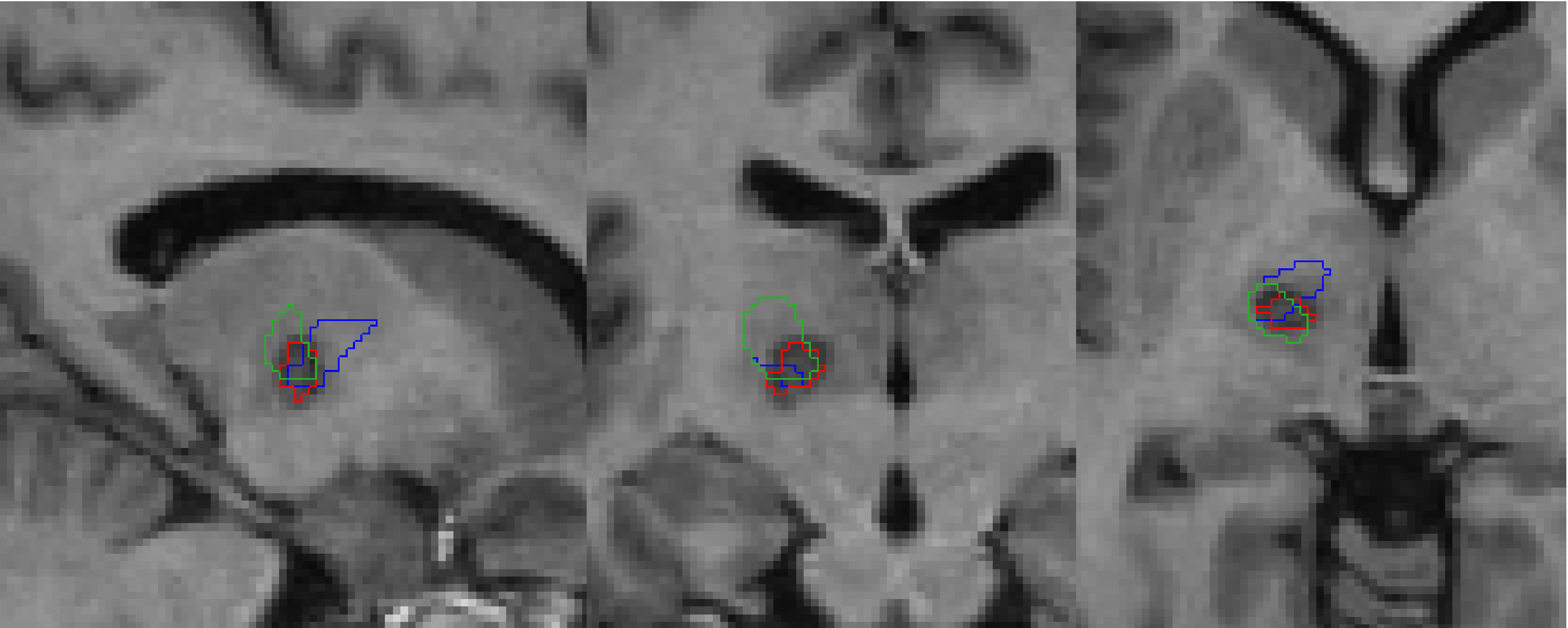}\\
    \includegraphics[width=10.0cm]{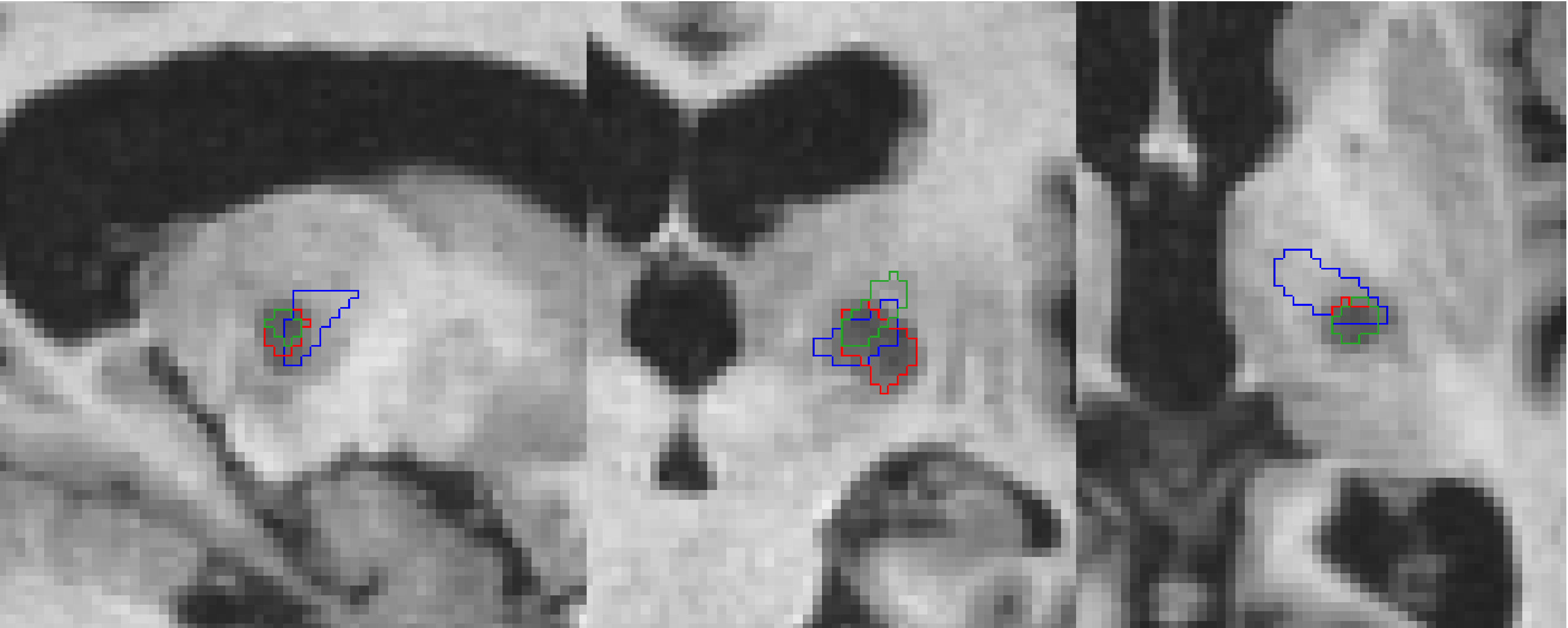}\\
    \includegraphics[width=10.0cm]{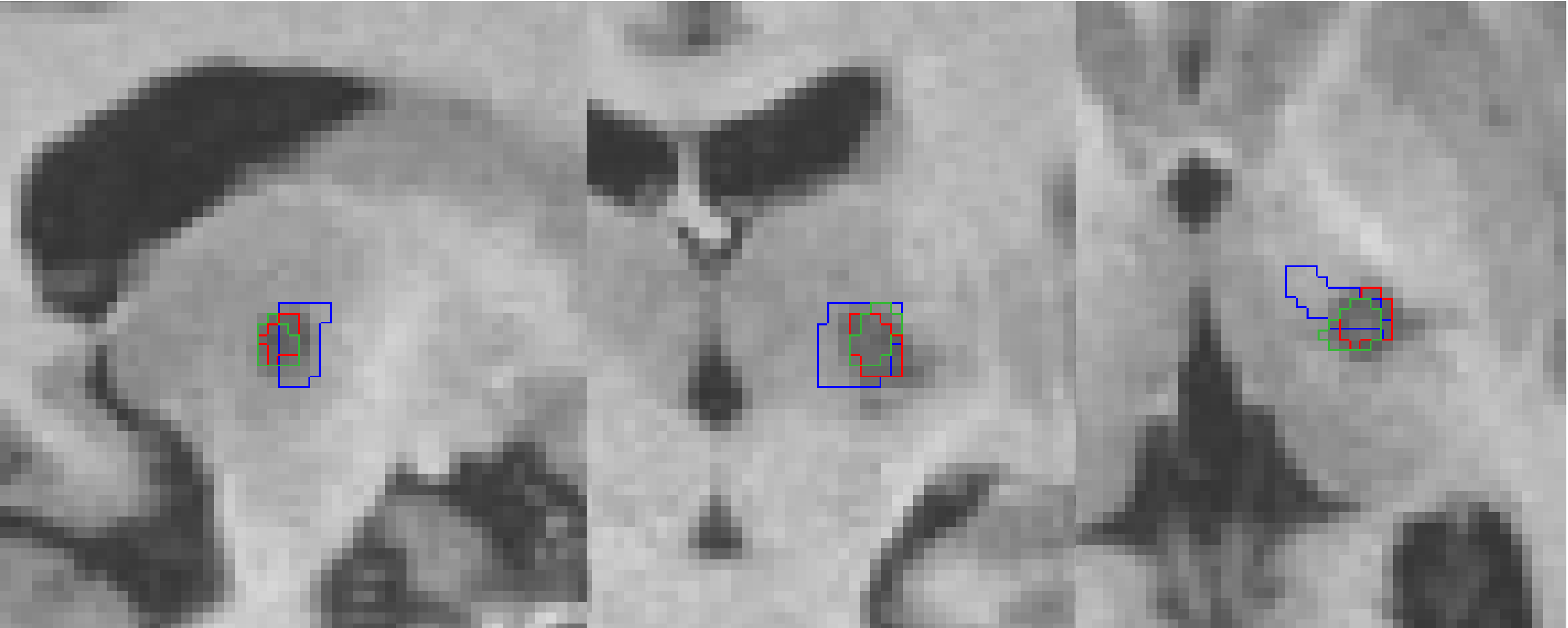}\\
     
     \caption{Comparison of VIM ROIs for three representative subjects. From left to right: sagittal, coronal and axial projection. Green: predicted ROI; red: reference ROI. Blue: THOMAS prediction. Subject (a): NN $DSC = 0.26$, $sDSC = 0.23$, $TRP = 0.53$, $Precision = 0.17$; THOMAS $DSC = 0.18$, $sDSC = 0.21$, $TRP = 0.45$, $Precision = 0.12$. 
     Subject (b): NN $DSC = 0.30$, $sDSC = 0.32$, $TRP = 0.25$, $Precision = 0.38$; THOMAS $DSC = 0.18$, $sDSC = 0.15$, $TRP = 0.30$, $Precision = 0.13$.
     Subject (c): NN $DSC = 0.42$, $sDSC = 0.38$, $TRP = 0.51$, $Precision = 0.36$; THOMAS $DSC = 0.31$, $sDSC = 0.27$, $TRP = 0.58$, $Precision = 0.21$. Note that patient a) was treated on right thalamus. The NN was trained on the left side only. However, thanks to data augmentation applied during the training, the NN is able to detect the VIM in the right side with similar performance as for patients b) and c).}  
     \label{fig_CLIN}
\end{figure}

Figure \ref{fig_CLIN} illustrates the spatial correspondence between the reference ROI (red contour), the NN prediction (green contour), and the THOMAS prediction (blue contour) for three representative subjects, displayed in the sagittal, coronal, and axial planes.

\begin{figure}[h!]
    \centering
     \begin{tabular}{cc}
    \includegraphics[width=6.4cm]{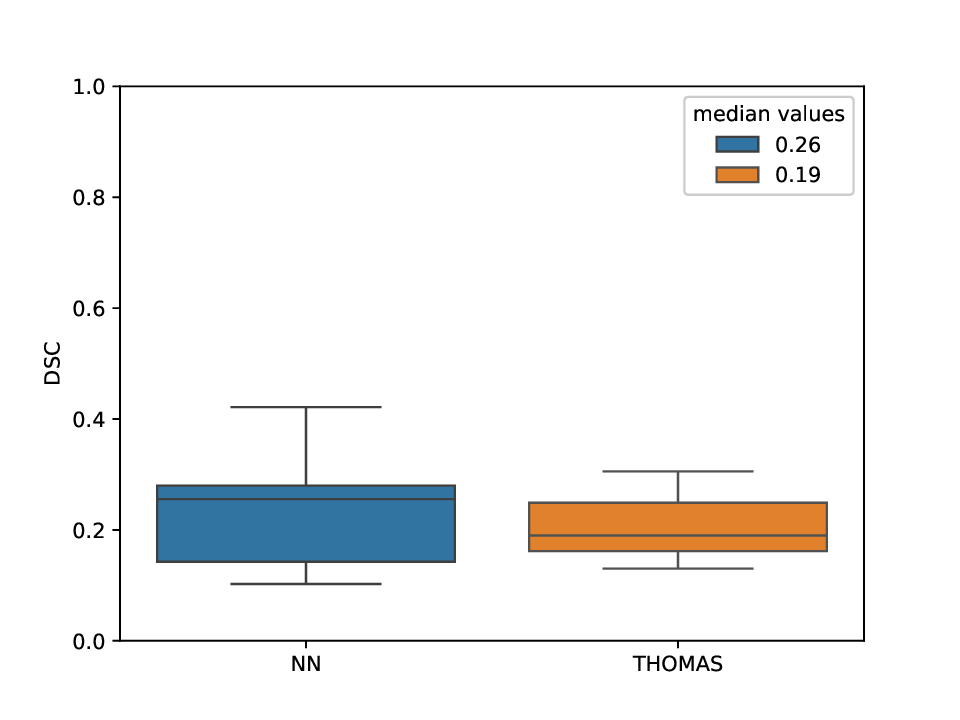} &
    \includegraphics[width=6.4cm]{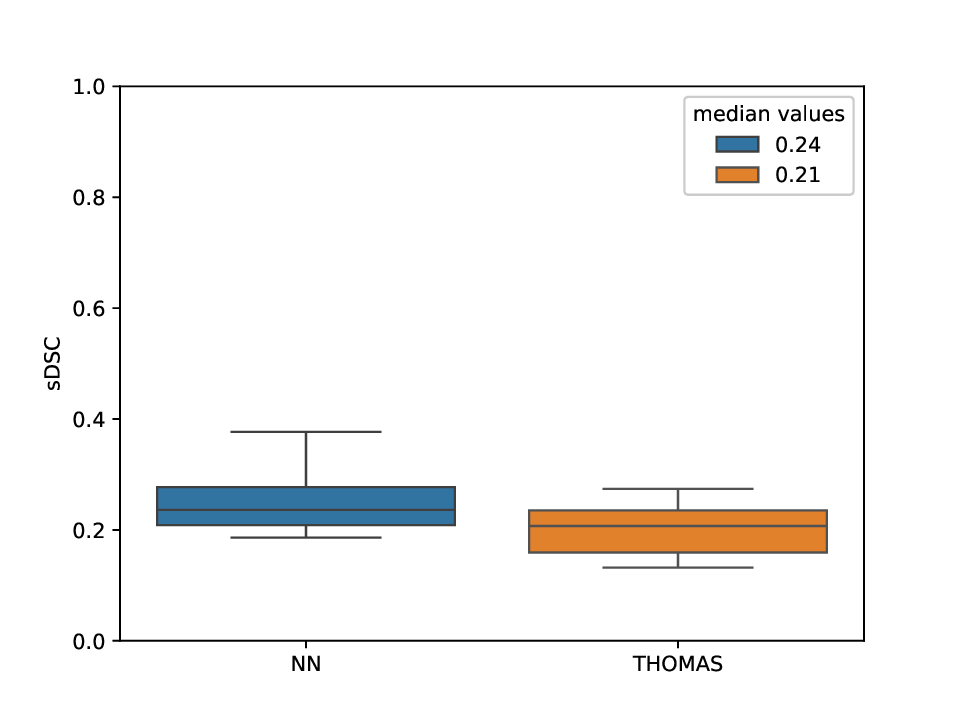} \\
    (a) & (b) \\    
    \includegraphics[width=6.5cm]{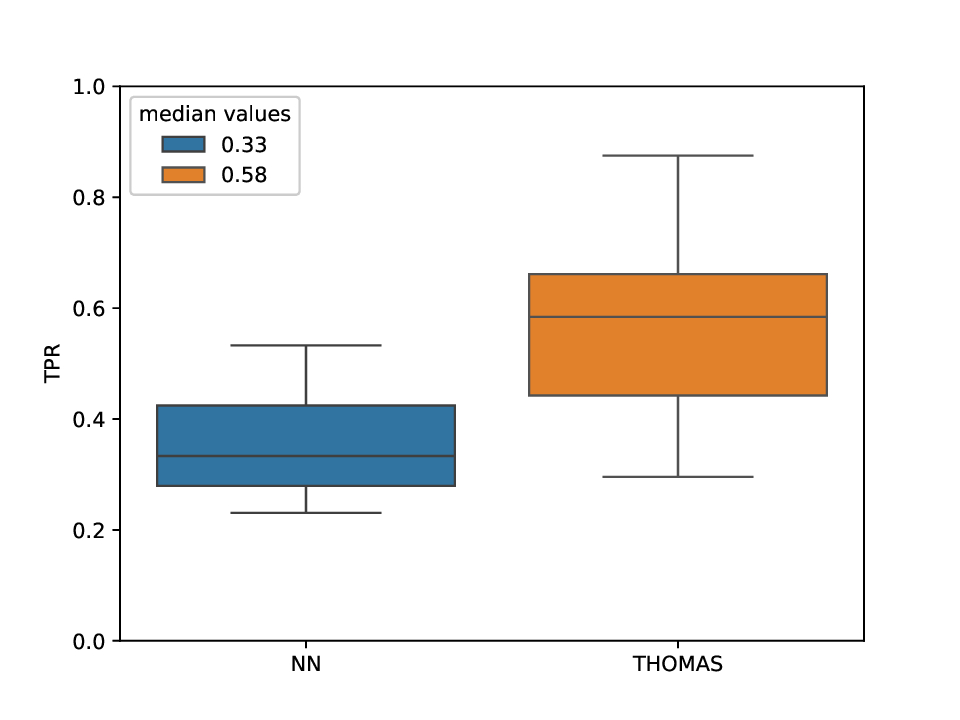} & \includegraphics[width=6.5cm]{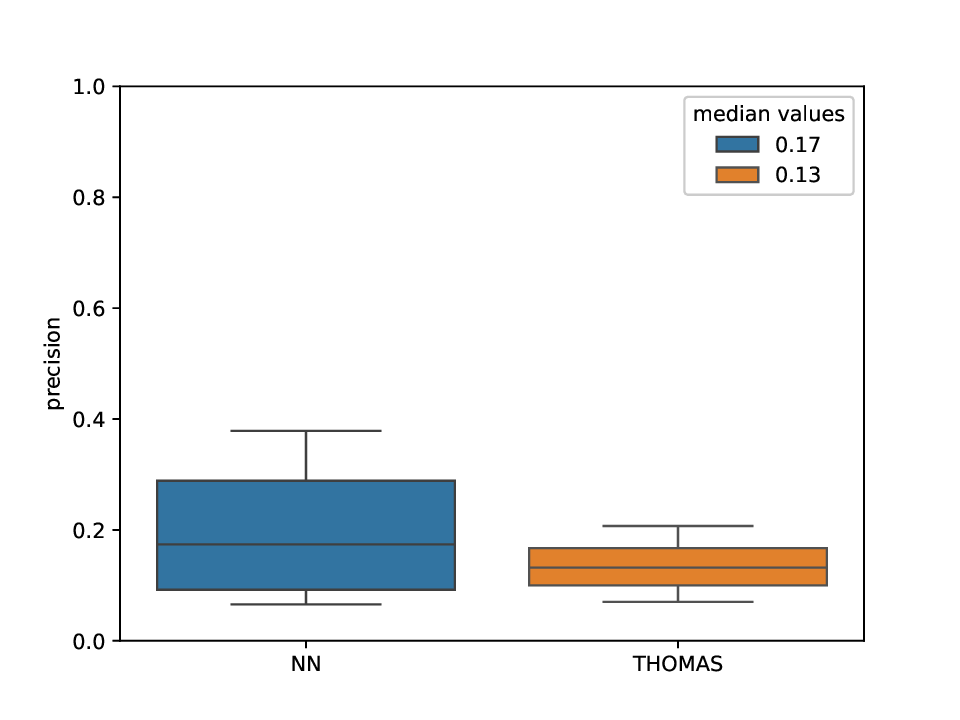} \\
    (c) & (d)
    \end{tabular}
    \caption{Metrics comparison between NN and THOMAS prediction ability. a) DSC distribution; b) sDSC distribution; c) TPR distribution; d) Precision.}
    \label{fig_boxPlot}
\end{figure}

Table \ref{tab_result2} and Figures \ref{fig_boxPlot} report the results for both the NN and THOMAS pipelines. Overall, the two methods exhibit comparable performance: NN attains marginally higher DSC and sDSC values than THOMAS (Figure \ref{fig_boxPlot}a,b). In terms of $TPR$, which represents the portion of GT volume recovered by the predictions, THOMAS performs better than NN (Figure \ref{fig_boxPlot}c). It is noteworthy, however, that THOMAS produces ROIs with an average volume of $371\ \text{mm}^3$, substantially larger than the $205\ \text{mm}^3$ predicted on average by NN (in this clinical dataset, the average $GT$ volume is $102\ \text{mm}^3$). This discrepancy suggests that the superior performance of THOMAS, in terms of $TPR$, may be influenced by a volume-related effect (see section \ref{evaluation}), as further supported by the observed $Precision$ values (Figure \ref{fig_boxPlot}d).

\begin{table}[h!]
    \centering

    \begin{tabular}{|c|c|c|c|c|c|}
        \hline
          method  & DSC & sDCS& TPR & Precision & D$_{COM}[mm]$\\
        \hline
         NN & $0.23\pm 0.11$ & $0.25\pm0.07$ & $0.36\pm0.12$ & $0.20\pm0.13$ & $4.5\pm1.5$\\
         \hline
         THOMAS & $0.21\pm0.07$ & $0.20\pm0.05$ & $0.57\pm0.19$ & $0.13\pm0.05$ & $4.0\pm1.1$\\
         \hline
    \end{tabular}
    \caption{Evaluation analysis with different metrics utilized.}
    \label{tab_result2}
\end{table}

\section{Discussion}
A deep learning approach has been developed to help targeting the ventral intermediate nucleus (VIM) in neurosurgical treatments of neurofunctional disorders as ET. This is achieved by predicting the VIM location using a CNN trained and validated on subject specific probabilistic tractography data, by exploiting structural (T$_1$-weighted images) and diffusion-weighted acquisition of more that 1,000 subjects of the HCP dataset. An additional validation was performed of real-world clinical images from a small sample of patients who underwent tcMRgFUS thalamotomy.\\
Recent literature reports that tractography-based targeting has important benefit for tcMRgFUS and DBS treatment planning and treatment \cite{feltrin2022focused,akram2018connectivity}. First, this is a personalized method as it takes into account the individual variability rather than using atlas-based indirect targeting, which, although standardized, has evident weakness. The choice of targeting through probabilistic tractography approach allows to consider the inherent variability in brain anatomy across patients and to improve outcomes of thalamotomy \cite{feltrin2022focused,tsolaki2018potential}. The probabilistic tractography approach shows reduced uncertainty in dense areas with crossing fibers and is less prone to sampling limitations with respect to deterministic tractography.
However, probabilistic approach could be computational demanding and time consuming \cite{behrens2007probabilistic,tsolaki2018potential}.
Indeed, since DTI images are prone to be affected by noise and motion artifacts, a pre-processing of diffusion data is mandatory. Then, BEDPOSTX and PROBTRAKX are very time consuming. In recent years GPU versions of these tools have been developed to significantly accelerate processing; however, the time required for the entire protocol is still too long for a real-time application in clinical practice. Furthermore, powerful GPU-based architectures are not easily available to clinicians for this purpose.

The model developed here is able to predict ROIs with good agreement to those tractography-based (internal validation, DSC $0.62\pm0.15$, sDSC $0.76\pm0.17$). Moreover, it requires only T$_1$ images, overcoming not only the limitation associated with DTI image pre-processing but also the acquisition time itself. On the computational side, once the network is trained, the time required for model querying is still very low even on CPU architectures.

Some limitations have however to be considered.\\
First, the model is not provided with any information on the diffusion. The tractography data are fed to the model only at the training stage to identify the desired region, while at the querying stage the model produces an output based on T$_1$ images only. As it is well known, DL models are able to learn correlations between input and target, in our case a T$_1$ image and the corresponding tractography-based region. Given the size of the dataset utilized, one could be confident that these correlations are well achieved. \\
Secondly, most functional neurological disorders are more common in aged people in which the brain typically shows more evident signs of atrophy with larger subarchnoid and ventricular spaces. The enlargement of the third ventricle in particular may result in significant difference of thalamic targets location compared to AC-PC references. The HCP dataset contains images from young adult subjects only, acquired at 3T, which introduces a bias in our trained model. This issue has been taken into account during the training by data augmentation. \\
Moreover, the NN was trained in predicting tractography-based ROIs, i.e. regions that overlaps with the VIM location with high probability, even though they are not exactly the same. The protocol for VIM location defined in section \ref{VIM_def}, based on sectioning the DRTC track estimated as reported in literature \cite{akram2018connectivity}, is not error-free. In ET surgical treatments the target is provided by a direct feedback of the patient. We must consider that the VIM location is still an open issue \cite{parras2022role}. In this respect, our work is meant as a support to reduce the region to inspect.\\ 
All these issues become evident in the external validation. Indeed, the results on DSC and sDSC are considerably worse compared to those obtained on the HCP data. This is certainly due to the differences in data quality and patient anatomy variability. However, it must be considered that in the external validation we used as GT the segmented region of the edema induced by thermal ablation.  This has two implications. First, this ROI also does not correspond exactly to the VIM, although it includes it. Second, the ROI is extracted from the follow-up image and it is used as GT for the screening images after appropriate co-registration. Notwithstanding co-registration, follow-up and screening images are different; this introduces an uncertainty that cannot be avoided and affects the results.\\
All the factors mentioned above also affect the results obtained with the THOMAS pipeline. The Dice coefficients are slightly higher for NN than for THOMAS, but these differences are not large enough to clearly determine which pipeline performs better. However, from a timing perspective, NN is able to provide predictions in a significantly shorter computational time compared to THOMAS (a few seconds per patient versus tens of minutes), providing a clear advantage.

Our findings indicate that the proposed CNN is suitable for real-time clinical applications and represents a solid foundation for future investigations and further developments. \\
The methodology described here is designed to overcome timing constraints, providing useful real-time information. Moreover, unlike atlas-based approaches, the pipeline we developed is freely available for both inference and training. Additional refinements of the VIM region could also be incorporated for fine-tuning or re-training, thereby potentially enhancing the network’s predictive performance.

\section{Conclusions}

In this study, we present a novel end-to-end open-source DL framework for identifying the VIM, the target of neurosurgical treatments in patients with ET or other movement disorders characterized by tremor. This is achieved by using a CNN trained in finding the VIM binary masks given T$_1$ images as input. 
The developed model has been trained on the HCP database. Two validation steps have been carried out. On one hand, we test the model capability in predicting ROIs obtained by probabilistic tractography on the HCP data (internal validation). On the other hand (external validation), the model is tested in predicting the VIM location on clinical data, along with a comparison with an atlas-based method (THOMAS). \\
VIM predictions are achieved in a very short time (fraction of second per subject), making the proposed method promising for live targeting in MR-guided procedures, providing further fine tuning based on clinical data sets.

\section*{Acknowledgments}
Raw data were provided by the Human Connectome Project, WU-Minn Consortium (Principal Investigators: David Van Essen and Kamil Ugurbil; 1U54MH091657) funded by the 16 NIH Institutes and Centers that support the NIH Blueprint for Neuroscience Research; and by the McDonnell Center for Systems Neuroscience at Washington University.\\
This work was received funding from the Italian Ministry of Health Ricerca Finalizzata 2016-Giovani Ricercatori, call under grant agreement no. GR-2016-02364526, from the Italian National Institute of Nuclear Physics with the research project Artificial Intelligence in Medicine (next-AIM) and the Ministy of Research with the Project "SiciliAn MicronanOTecH Research And Innovation CEnter "SAMOTHRACE" (MUR, PNRR-M4C2, ECS\_00000022), spoke 5 - Università degli Studi di Palermo "S2-COMMs - Micro and Nanotechnologies for Smart \& Sustainable Communities".
Research partly supported by PNRR - M4C2 - Investimento 1.3, Partenariato Esteso PE00000013 - "FAIR - Future Artificial Intelligence Research" - Spoke 8 "Pervasive AI", funded by the European Commission under the NextGeneration EU programme.
This work was received funding also from the project "Italian network of excellence for advanced diagnosis (INNOVA) - Advanced Diagnostic Platform (HLS-DA)" funded by the Italian Ministry of Health as part of the National Complementary Plan for the Innovative Health Ecosystem (PNC-3-2022-23683266).\\
Mattia Romeo acknowledge funding from UPMC Italy srl.

\bibliographystyle{unsrt}
\bibliography{refs}

\end{document}